\documentclass[sigconf]{acmart}
\AtBeginDocument{%
  }

\copyrightyear{2026}
\acmYear{2026}
\setcopyright{cc}
\setcctype{by-nc-nd}
\acmConference[ICAIF '26]{7th ACM International Conference on AI in Finance}{November 14--17, 2026}{Milan, AA, Italy}
\acmBooktitle{7th ACM International Conference on AI in Finance (ICAIF '26), November 14--17, 2026, Milan, AA, Italy}
\acmDOI{10.1145/3858063.3858823}
\acmISBN{979-8-4007-3087-0/2026/11}

\usepackage{pifont}     % for \ding tick and cross
\newcommand{\cmark}{\ding{51}}   % tick
\newcommand{\xmark}{\ding{55}}   % cross
\usepackage{makecell}   
\usepackage{enumitem}
\usepackage{subcaption} 
\usepackage{amsmath}
\usepackage{tikz}
\usepackage{booktabs}

\usepackage{etoolbox}

\newcommand{\ie}{\emph{i.e., }}
\newcommand{\eg}{\emph{e.g., }}

\newcommand{\etc}{\emph{etc.}}

\newcommand{\ow}{\mathrm{ow}}
\newcommand{\act}{\mathrm{act}}
\newcommand{\fund}{\mathrm{fund}}
\newcommand{\tech}{\mathrm{tech}}

\usetikzlibrary{positioning, arrows.meta, fit, backgrounds, calc,
    decorations.pathreplacing, shadows.blur, shapes.geometric}
\pgfdeclarelayer{deepbg}
\pgfsetlayers{deepbg,background,main}
\providecommand{\footref}[1]{}

\begin{document}

\title{Regime-Aware Interpretable Ranking for Sector
Rotation}

\author{Kelvin J.L. Koa}
\affiliation{%
  \institution{Asian Institute of Digital Finance,
               National University of Singapore}
  \country{Singapore}
  }
\email{kelvin.koa@u.nus.edu}

\author{Ziling Liu}
\authornote{These authors contributed equally to this research.}
\affiliation{%
  \institution{Asian Institute of Digital Finance,
               National University of Singapore}
  \country{Singapore}
  }
\email{e1547130@u.nus.edu}

\author{Xinyang Li}
\authornotemark[1]
\affiliation{%
  \institution{Asian Institute of Digital Finance,
               National University of Singapore}
  \country{Singapore}
  }
\email{e1561242@u.nus.edu}

\author{Shuowei Feng}
\authornotemark[1]
\affiliation{%
  \institution{Asian Institute of Digital Finance,
               National University of Singapore}
  \country{Singapore}
  }
\email{e1539221@u.nus.edu}

\author{Ruoyu Liang}
\authornotemark[1]
\affiliation{%
  \institution{Asian Institute of Digital Finance,
               National University of Singapore}
  \country{Singapore}
  }
\email{e1576578@u.nus.edu}

\author{Mathieu Giroux}
\authornote{The views and opinions expressed in this article are
solely those of the authors and do not necessarily reflect the
views, positions, or policies of Manulife or any of its affiliates.
Any methodological or empirical errors are the sole responsibility
of the authors and should not be attributed to Manulife.}
\affiliation{%
  \institution{Manulife Investment Management}
  \city{Montreal}
  \state{Quebec}
  \country{Canada}}
\email{mathieu\_giroux@manulifeam.com}

\author{Eugene Y. Wen}
\authornotemark[2]
\affiliation{%
  \institution{Manulife}
  \city{Toronto}
  \state{Ontario}
  \country{Canada}}
\email{eugene\_wen@manulife.com}

\author{Xuan Yao}
\affiliation{%
  \institution{Asian Institute of Digital Finance,
               National University of Singapore}
  \country{Singapore}
  }
\email{yaoxuan@nus.edu.sg}

\author{Ke-Wei Huang}
\affiliation{%
  \institution{Asian Institute of Digital Finance,
               National University of Singapore}
  \country{Singapore}
  }
\email{dishkw@nus.edu.sg}

\author{Hugo Lamarre}
\authornotemark[2]
\affiliation{%
  \institution{Manulife Investment Management}
  \city{Montreal}
  \state{Quebec}
  \country{Canada}}
\email{hugo\_lamarre@manulifeam.com}

%%
%% Concise author list for page headers.
\renewcommand{\shortauthors}{Koa et al.}



\begin{abstract}
Sector rotation is a common decision in benchmark-relative portfolio management, where portfolio managers rank sectors for overweight and underweight positions. We formulate this task as ranking sector actions by expected future risk-adjusted active performance, measured with forward sector--benchmark Sharpe ratios. Such rankings should preserve temporal information, express model outputs through economically familiar signal characteristics, account for changes in signal relevance across macroeconomic environments, and capture dependencies among sectors. We propose \mbox{\textsc{RegimeRank}}, an interpretable sector-ranking model that represents fundamental and technical signals through expert-defined temporal characteristics, conditions them on learned macro factors via cross-attention, and models cross-sector dependencies through self-attention. Expected-gradient attributions identify the contributions of individual signal--transformation pairs, while macroeconomic and cross-sector attention maps provide diagnostics of the learned conditioning and relational structure. We evaluate \mbox{\textsc{RegimeRank}} on weekly S\&P 500 data from 2002 to 2024 under combinatorial purged cross-validation, where it outperforms representative ranking and machine-learning baselines across multiple ranking metrics. Ablation studies support the contributions of latent macro states, regime conditioning, and cross-sector modelling, while attribution and attention analyses illustrate the temporal characteristics, macro states, and sector relationships underlying its rankings.
% \vspace{-5px}
\end{abstract}

% \vspace{-10px}

% \ccsdesc[300]{Applied computing~Forecasting}


% \vspace{-5px}

\maketitle
% \vspace{-4px}

% \begingroup
% \renewcommand{\thefootnote}{}%
% \footnotetext{\textit{Disclaimer.} This article is presented for research and
% informational purposes only and does not constitute investment advice or a
% recommendation to buy or sell any security or financial instrument. The results
% presented are based on simulated performance; simulated results have inherent
% limitations, and neither simulated nor past performance is indicative of future
% results.}%
% \endgroup

\section{Introduction}

\begin{table*}[t]
% \vspace{-5px}
\centering
\small
\captionsetup{skip=2pt}
\caption{Comparison of methodological capabilities across financial and
AI-based asset-ranking approaches.}
\label{tab:comparison}
\renewcommand{\arraystretch}{0.88}
\begin{tabular}{l cccccc}
\toprule
\textbf{Approach}
& \makecell{Ranking\\Formulation}
& \makecell{Temporal\\Dynamics}
& \makecell{Regime\\Aware}
& \makecell{Asset\\Relational}
& \makecell{Mechanism\\Interpretability}
& \makecell{Attribution\\Interpretability} \\
\midrule
\textbf{Asset Ranking in Finance} & & & & & & \\
RankNet / ListNet \cite{song2017stock}
& \cmark & \xmark & \xmark & \xmark & \cmark & \xmark \\
Cross-sectional LTR \cite{poh2020building}
& \cmark & \xmark & \xmark & \xmark & \cmark & \xmark \\
Listwise Node2Vec \cite{saha2021stock}
& \cmark & \cmark & \xmark & \cmark & \cmark & \xmark \\
Long-Short Listwise LTR \cite{zhang2022constructing}
& \cmark & \xmark & \xmark & \xmark & \cmark & \xmark \\
SABER \cite{kouloumpris2024saber}
& \cmark & \xmark & \xmark & \xmark & \cmark & \xmark \\
\midrule
\textbf{Asset Ranking in AI} & & & & & & \\
RSR \cite{feng2019temporal}
& \cmark & \cmark & \xmark & \cmark & \xmark & \xmark \\
STHAN-SR \cite{sawhney2021stock}
& \cmark & \cmark & \xmark & \cmark & \xmark & \xmark \\
RT-GCN \cite{zheng2023relational}
& \cmark & \cmark & \xmark & \cmark & \xmark & \xmark \\
ADB-TRM \cite{chen2024automatic}
& \cmark & \cmark & \cmark & \cmark & \xmark & \xmark \\
PDU \cite{huang2025progressive}
& \cmark & \cmark & \xmark & \cmark & \xmark & \xmark \\
\midrule
\textsc{RegimeRank} (Ours)
& \cmark & \cmark & \cmark & \cmark & \cmark & \cmark \\
\bottomrule
\end{tabular}
\end{table*}

Sector rotation is a core decision in active portfolio
management, in which portfolio managers (PMs) determine which sectors to
overweight as economic conditions evolve
\cite{moskowitz1999industries,o2000industry}. The economic significance of
this decision is reflected in the persistent dispersion of sector returns:
between 1989 and 2025, the average calendar-year difference between the
best- and worst-performing S\&P 500 sectors was approximately 45 percentage
points \cite{pathak2026sector}. Such dispersion creates both an opportunity
and a risk for benchmarked portfolios: overweighting the right sectors can
enhance active performance.

\begingroup
\renewcommand{\thefootnote}{}%
% \vspace{-10px}
\footnotetext{\textit{Disclaimer.} This article is presented for research and
informational purposes only and does not constitute investment advice or a
recommendation to buy or sell any security or financial instrument. The results
presented are based on simulated performance; simulated results have inherent
limitations, and neither simulated nor past performance is indicative of future
results.}%
\endgroup

The central challenge is not to identify sector winners and losers
ex post, but to rank their relative attractiveness ex ante. Because portfolio
tilts naturally emerge from an ordering of sectors, we formulate sector
rotation as a learning-to-rank problem
\cite{alsulmi2022ranking,poh2020building,saha2021stock,
zhang2022constructing}. This formulation avoids requiring precise return
forecasts---a particularly difficult task in financial markets
\cite{feng2019temporal,kouloumpris2024saber}---and instead focuses the model
on the relative ordering that directly informs overweight and underweight
decisions. We rank portfolios on forward Sharpe ratios to ensure that a sector is rewarded for return earned \emph{per unit of risk}. This is consistent with benchmarked mandates to surface best risk-adjusted opportunities. Our model combines fundamental and technical signals, conditioning both on macroeconomic indicators that characterize the prevailing market environment.

Predictive accuracy alone, however, is insufficient. PMs remain accountable to investment and risk committees and must therefore
be able to understand and communicate the rationale behind a
recommended sector tilt. We consequently target two complementary forms of
interpretability. \emph{Mechanism interpretability} requires the model's
internal components to correspond to economically meaningful operations,
whereas \emph{attribution interpretability} requires an individual ranking
outcome to be traced to the signals that drive it. Existing sequential models
make this alignment difficult because they typically encode historical
observations into latent representations. Post-hoc methods such as TimeSHAP
\cite{bento2021timeshap} can attribute a prediction to particular time steps,
but PMs more commonly reason in terms of higher-level characteristics
such as trend, persistence, and volatility. An interpretable sector
ranking model should preserve temporal information while expressing
it through concepts that can be related to investment reasoning.

\begin{figure*}[t]
\centering
% \vspace{-5px}
\includegraphics[width=0.98\linewidth]{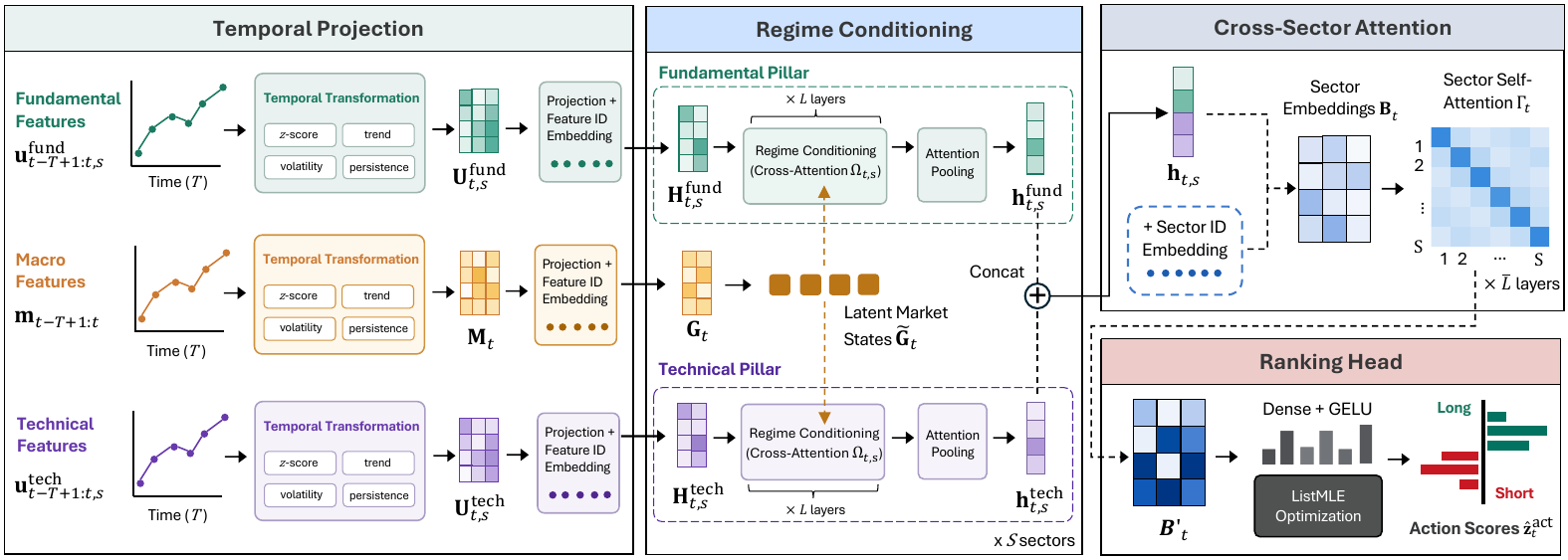}
% \input{fig_framework}
% \vspace{-17px}
\caption{
Overall framework for \mbox{\textsc{RegimeRank}}. The solid \textit{Regime Conditioning} box is the per-sector plate, replicated $S$ times, and the dashed boxes are the two \textit{per-pillar encoders}, conditioned on the shared latent states $\widetilde{\mathbf{G}}_{t}$ (dashed arrows). Boxes are labelled by their outputs, and the pillar embeddings are concatenated before the \textit{Cross-Sector Attention} component.
}
% \vspace{-7px}
\label{fig:framework}
\end{figure*}

A common sector-allocation workflow suggests three additional modeling
requirements. First, PMs form complementary views from fundamental signals,
such as valuations, earnings, and growth, and technical
signals, such as momentum, volume, and volatility, before combining them into
an overall conviction. Second, the relevance of these signals may vary across
macroeconomic environments: a low valuation, for example, may indicate an
oversold investment opportunity in one environment but deteriorating fundamentals in another.
Third, sectors cannot be evaluated independently. Sectors exposed to common inflation, commodity, or interest-rate forces often move jointly, so
information about one sector may affect the relative attractiveness of
others. These considerations motivate a model that separately represents
fundamental and technical information, conditions their interpretation on the
macroeconomic environment, and jointly models cross-sector dependencies.

We propose \textsc{RegimeRank}, an interpretable learning-to-rank architecture with four components. First, a \emph{temporal-projection} layer
represents each signal through expert-defined characteristics \cite{cont2001empirical,carbone2004time}. Second, a
\emph{macro-conditioning} module uses cross-attention to condition these
signal representations on learned macroeconomic factors, allowing their
relevance to vary across market environments
\cite{chen2024deep,pelger2022state}. Temporal projection and macro
conditioning are applied separately within fundamental and technical
encoders, preserving the distinction between the two information pillars. Third, a \emph{cross-sector
self-attention} layer then propagates information across sectors exposed to
related economic forces
\cite{feng2019temporal,sawhney2021stock,poh2021enhancing,
lopez2016building}. Finally, a listwise ranking head
\cite{xia2008listwise} converts the resulting sector representations into an
ordering of overweight and underweight opportunities.

Through extensive experiments, we show that \textsc{RegimeRank} outperforms
representative ranking, machine-learning and deep-learning baselines across
multiple evaluation metrics in a simulation study applied to the S\&P 500 sectors. Ablation studies demonstrate the contribution of
each architectural component, while analyses of Shapley values
\cite{lundberg2017unified} and attention weights \cite{bahdanau2014neural,
clark2019does} provide insights into the temporal characteristics, market
regimes, and sector interdependencies underlying its rankings.
% \vspace{-3px}
\section{Related Work}

\subsection{Asset Ranking}
Asset ranking has emerged as an active research area across both the
finance and artificial intelligence (AI) literature. In both
communities, the formulation is motivated by the relative nature of
investment decisions \cite{saha2021stock}: investors primarily need
to identify relative winners, a target that generalizes
better and is more robust than predicting exact future returns
\cite{poh2020building, feng2019temporal}. Finance works build upon
this formulation by incorporating financial structure, such as news
sentiment \cite{song2017stock}, long-short strategies
\cite{zhang2022constructing} and price stochasticity
\cite{kouloumpris2024saber} into the ranking optimization. AI works
instead extend the learning machinery, through graph convolutions
\cite{zheng2023relational}, adversarial training
\cite{chen2024automatic} and contrastive learning
\cite{huang2025progressive}. As summarized in Table~\ref{tab:comparison}, few surveyed methods combine temporal or regime structure with mechanisms a PM could interpret, and we are not aware of existing approaches that provide attribution over their ranking outcomes.

% \vspace{-3px}
\subsection{Model Interpretability}
Interpretability in financial models \cite{shmueli2010explain, koa2026reasoning, koa2026selfevolving} has been
approached from two complementary directions. Mechanism interpretability
seeks economically meaningful explanations of model behavior, with examples
ranging from nonlinear factor interpretations of neural networks
\cite{kelly2024large} to attention-based asset-pricing models
\cite{kelly2025artificial} and graph-based dependency modeling
\cite{sarmah2024learning}. Attribution interpretability, in contrast, focuses
on identifying the signals responsible for model predictions through methods
such as SHAP \cite{lundberg2017unified}, LIME \cite{ribeiro2016should},
TreeInterpreter \cite{sharma2020evaluating}, Accumulated Local Effects
\cite{apley2020visualizing}, and attention-based explanations
\cite{clark2019does, mannix2024comfe}.
% \vspace{-2px}
\section{The \textsc{RegimeRank} Framework}

\subsection{Architecture Overview}
The goal is to rank sector overweights and underweights relative to a
benchmark. \textsc{RegimeRank} (Figure~\ref{fig:framework}) proceeds in four stages:
\begin{enumerate}[nosep,leftmargin=*]
\item \textbf{Temporal Projection.} Encode each feature's recent
history into a fixed-length temporal summary (trend, volatility, \etc).
\item \textbf{Regime conditioning.} Let sector signals attend to macro
indicators so that the same valuation or momentum reading can mean
different things under different economic regimes.
\item \textbf{Cross-sector attention.} Allow sectors to exchange information;
\eg a rally in cyclicals informs the outlook for defensives.
\item \textbf{Ranking head.} Map the resulting sector embeddings to scalar
scores whose ordering defines the recommended tilt ranking.
\end{enumerate}
Stages 1--2 run independently for each information pillar (fundamentals and
technicals); the per-pillar embeddings are concatenated before stage~3. This
separation reflects how practitioners typically form a fundamental view and a
technical view before combining them into a single conviction.

%%%%%%%%%%%%%%%%%%%%%%%%%%%
% \vspace{-2px}
\subsection{Notation Conventions}
Lowercase bold denotes vectors ($\mathbf{u}$), uppercase bold denotes
matrices ($\mathbf{W}$), and uppercase italics denote dimensions ($S$, $F$,
$N$, $D$).
Subscripts are reserved for integer indices
(\eg\ $t$ for time steps, $s$ for sectors).
Superscripts identify the block or parameter family
(\eg\ $\mathbf{W}^Q$, $\mathbf{h}^{\fund}$).
A hat marks predictions ($\hat{\mathbf{z}}$).
Parentheses index vector entries ($y(a)$ is the $a$-th entry of
$\mathbf{y}$), and primes mark successive residual updates within a block
($\mathbf{H}$, $\mathbf{H}'$, $\mathbf{H}''$).

%%%%%%%%%%%%%%%%%%%%%%%%%%%
\subsection{Problem Setting}
For each sector $s \in \{1,\ldots,S\}$ at timestamp $t$, we observe a
\textit{sector-specific} feature vector $\mathbf{u}_{t,s} \in
\mathbb{R}^F$ (fundamental and technical signals, processed in separate
pillars) and a \textit{shared} macro vector $\mathbf{m}_t \in \mathbb{R}^N$.
Each action corresponds to overweighting \emph{or} underweighting a sector against
the benchmark; we index actions by
$a \in \{1,\ldots,2S\}$. Let $\mathbf{y}^{\ow}_t \in \mathbb{R}^S$
denote the vector of \emph{forward Sharpe ratios}
calculated on the sector returns in excess of the benchmark return\footnote{Computed ex post: the
mean of spread returns over the forward window divided by their standard
deviation over the same window. Risk-free rate is set to zero since spreads
are self-financed. This quantity is also known as the \emph{information ratio}.}.
The action-level target stacks overweight and underweight sides:
\[
\mathbf{y}^{\act}_t
=
\begin{bmatrix}
\mathbf{y}^{\ow}_t \\[2pt]
-\mathbf{y}^{\ow}_t
\end{bmatrix}
\in \mathbb{R}^{2S}.
\]
The learning target is the permutation $\pi_t$ that sorts actions from best
to worst by realized quality: $\pi_t(1)$ is the index of the highest-valued
action, $\pi_t(2S)$ the lowest. Formally,
$y^{\act}_t(\pi_t(1)) \ge y^{\act}_t(\pi_t(2)) \ge \cdots
\ge y^{\act}_t(\pi_t(2S))$.
The model produces scores $\hat{\mathbf{z}}^{\act}_t \in \mathbb{R}^{2S}$ whose
induced ordering is trained to match $\pi_t$ via a listwise likelihood. The
supervision is purely ordinal ($\hat{\mathbf{z}}^{\act}_t$ is identified only up
to order-preserving transformations), consistent with the motivation that
relative rankings are more robust than exact performance levels.

%%%%%%%%%%%%%%%%%%%%%%%%%%%
\subsection{Temporal Projection}
The temporal projection, regime conditioning, and attention pooling stages
are applied independently to the fundamental and technical pillars. We
describe a single pillar and suppress the pillar tag until the merge; the
other pillar is identical. Within a pillar, $F$ denotes that pillar's
feature count.

The model ingests a cross-sectional snapshot, so temporal dynamics must be
encoded explicitly. Each feature's values over a lookback window of length
$T$ are mapped by a transformation operator $\mathcal{T}$ to a
$P$-dimensional vector of temporal summaries
($z$-scores, trend, volatility, persistence). For sector-specific features:
\[
\mathbf{U}_{t,s}
=
\mathcal{T}\!\left(
\mathbf{u}_{t-T+1:t,s}
\right)
\in \mathbb{R}^{F\times P},
\]
where each row holds the $P$ transformed values of one feature. 

These are projected onto a $D$-dimensional hidden space:
\[
\mathbf{H}_{t,s}
=
\mathbf{U}_{t,s}\mathbf{W}^{U}
+
\mathbf{E}^{U}
\in \mathbb{R}^{F\times D},
\]
where $\mathbf{W}^{U}\in\mathbb{R}^{P\times D}$ is the learned projection
matrix and $\mathbf{E}^{U}\in\mathbb{R}^{F\times D}$ contains learned
feature-identity embeddings that disambiguate features with otherwise
identical transformed values. The shared macro features are processed
analogously:
\[
\mathbf{M}_t
=
\mathcal{T}\!\left(
\mathbf{m}_{t-T+1:t}
\right)
\in \mathbb{R}^{N\times P},
\qquad
\mathbf{G}_{t}
=
\mathbf{M}_t\mathbf{W}^{M}
+
\mathbf{E}^{M}
\in \mathbb{R}^{N\times D},
\]
with $\mathbf{W}^{M}\in\mathbb{R}^{P\times D}$ and
$\mathbf{E}^{M}\in\mathbb{R}^{N\times D}$.

\subsubsection{Transformation-level Attribution}
For interpretability, we compute approximate SHAP values via expected
gradients \cite{erion2021improving} on the transformed inputs
\textit{before} projection. Let $\mathbf{X}_{t,s} = (\mathbf{U}_{t,s},
\mathbf{M}_t)$ collect the transformations for sector $s$. The attribution
is:
\[
\boldsymbol{\Phi}_{t,a,s} \approx
\mathbb{E}_{\mathbf{X}',\,\beta\sim\mathcal{U}(0,1)} \left[
\left(\mathbf{X}_{t,s}-\mathbf{X}'\right) \odot
\nabla_{\mathbf{X}_{t,s}} \hat{z}^{\act}_t(a) \Big|_{\mathbf{X}'
+\beta(\mathbf{X}_{t,s}-\mathbf{X}')} \right],
\]
where $\mathbf{X}'$ is a background observation and
$\beta\sim\mathcal{U}(0,1)$. The values approximate how much each feature
transformation contributes to the action score relative to the background
distribution.

%%%%%%%%%%%%%%%%%%%%%%%%%%%
\subsection{Regime Conditioning}
The informational value of sector signals is regime-dependent \cite{remillard2017option}: the same
signal may be bullish in one regime and bearish in another. We reduce the
macro features to $R$ latent market states:
\[
\widetilde{\mathbf{G}}_{t}
=
\mathbf{W}^{R}\mathbf{G}_{t}
\in \mathbb{R}^{R\times D},
\]
where $\mathbf{W}^{R}\in\mathbb{R}^{R\times N}$ mixes across the
\textit{feature} axis (not the hidden axis); each row represents a latent
aspect of the prevailing regime. In practice, the projection and reduction
are factored into two stages with an intermediate normalization.

We then condition sector features on these states via multi-head cross-attention,
using the sector features as queries and the latent macro states as keys
and values. We define:
\begin{equation*}
\mathbf{Q}_{t,s} = \mathbf{H}_{t,s}\mathbf{W}^{Q}
   \in \mathbb{R}^{F\times d},
\qquad
\star_{t} = \widetilde{\mathbf{G}}_{t}\mathbf{W}^{\star}
   \in \mathbb{R}^{R\times d},
\;\;
\star\in\{K,V\},
\end{equation*}
where $\mathbf{W}^Q\in\mathbb{R}^{D\times d}$ and
$\mathbf{W}^K,\mathbf{W}^V
\in\mathbb{R}^{D\times d}$, and $d$ is the attention projection width.
The attention weights and residual update are computed as:
\[
\boldsymbol{\Omega}_{t,s}
=
\operatorname{softmax}\!\!\left(
\frac{
\mathbf{Q}_{t,s}
{\mathbf{K}_{t}}^{\!\top}
}
{\sqrt{d}}
\right)
\in\mathbb{R}^{F\times R}.
\]
The context is projected back via a residual connection:
\[
\mathbf{H}'_{t,s}
=
\mathbf{H}_{t,s}
+
\bigl(\boldsymbol{\Omega}_{t,s}
\mathbf{V}_{t}\bigr)
\mathbf{W}^O
\in\mathbb{R}^{F\times D},
\]
where $\mathbf{W}^O\in\mathbb{R}^{d\times D}$.
Following the block convention, a position-wise FFN (with GELU activation)
completes each layer:
\[
\mathbf{H}''_{t,s}
=
\mathbf{H}'_{t,s}
+
\operatorname{FFN}\!\left(
\operatorname{LayerNorm}\bigl(\mathbf{H}'_{t,s}\bigr)
\right).
\]
Multi-head attention with $L$ stacked layers is used, each layer
recomputing $\boldsymbol{\Omega}_{t,s}$ from its input.
The attention map
$\boldsymbol{\Omega}_{t,s}$ is directly inspectable: it
shows how strongly the model conditions each signal on the prevailing macro
environment; cheap valuations may signal opportunity in one regime and a
value trap in another.

\subsubsection{Attention Pooling and Pillar Merge}
The regime-conditioned features are aggregated into an  embedding via
attention pooling:
\[
\boldsymbol{\alpha}_{t,s}
=
\operatorname{softmax}
\left(
\mathbf{H}''_{t,s}\mathbf{w}^{p} + b^{p}
\right)
\in\mathbb{R}^{F},
\qquad
\mathbf{h}_{t,s}
=
{\mathbf{H}''_{t,s}}^{\top}\boldsymbol{\alpha}_{t,s}
\in\mathbb{R}^{D},
\]
where $\mathbf{w}^{p}\in\mathbb{R}^{D}$ and
$b^{p}\in\mathbb{R}$ are learned pooling parameters. The weights
$\boldsymbol{\alpha}_{t,s}$ are inspectable and indicate the relative
contribution of each feature under the current regime.

This encoder runs independently for each pillar. The downstream sector
embedding concatenates the pillar embeddings:
\[
\mathbf{h}_{t,s}
=
\left[
\mathbf{h}_{t,s}^{\fund}\,;\,
\mathbf{h}_{t,s}^{\tech}
\right]
\in\mathbb{R}^{D},
\]
where $D=D^{\fund}+D^{\tech}$ denotes the
combined dimension.

%%%%%%%%%%%%%%%%%%%%%%%%%%%
\subsection{Cross-Sector Attention}
Sector rankings are shaped by interdependencies: cyclicals move together
and against defensives as the outlook shifts
\cite{saha2021stock, feng2019temporal}. To model this, we stack the sector
embeddings and add learned sector-identity embeddings (analogous to
positional encodings \cite{vaswani2017attention}):
\[
\mathbf{B}_t
=
\begin{bmatrix}
\mathbf{h}_{t,1}^\top\\
\vdots\\
\mathbf{h}_{t,S}^\top
\end{bmatrix}
+
\mathbf{E}
\in\mathbb{R}^{S\times D}.
\]
We define:
\begin{equation*}
\bar{\star}_{t} = \mathbf{B}_{t}\bar{\mathbf{W}}^{\star}
   \in \mathbb{R}^{S\times \bar{d}},
\qquad
\star\in\{Q,K,V\},
\end{equation*}
where $\bar{\mathbf{W}}^Q, \bar{\mathbf{W}}^K,
\bar{\mathbf{W}}^V
\in\mathbb{R}^{D\times \bar{d}}$, with projection width
$\bar{d}$.
Self-attention is applied:
\[
\boldsymbol{\Gamma}_t
=
\operatorname{softmax}\!\!\left(
\frac{
\bar{\mathbf{Q}}_t
{\bar{\mathbf{K}}_t}^{\!\top}
}
{\sqrt{\bar{d}}}
\right)
\in\mathbb{R}^{S\times S}.
\]
The context is projected and combined via a residual:
\[
\mathbf{B}'_t
=
\mathbf{B}_t
+
\bigl(\boldsymbol{\Gamma}_t
\bar{\mathbf{V}}_t\bigr)
\bar{\mathbf{W}}^O
\in\mathbb{R}^{S\times D},
\]
where $\bar{\mathbf{W}}^O
\in\mathbb{R}^{\bar{d}\times D}$.
The map $\boldsymbol{\Gamma}_t$ surfaces the cross-sector
relationships encoded by the network; a portfolio manager can compare these
against their own reading of the cross-section.

A position-wise FFN with GELU activation completes each block (following the
pre-norm convention):
\[
\mathbf{B}''_t
=
\mathbf{B}'_t
+
\operatorname{FFN}\!\left(
\operatorname{LayerNorm}(\mathbf{B}'_t)
\right).
\]
Multi-head attention and $\bar{L}$ layers, each recomputing
$\boldsymbol{\Gamma}_t$ from its input, produce the final
sector embeddings $\mathbf{B}''_t$.

%%%%%%%%%%%%%%%%%%%%%%%%%%%
\subsection{Ranking Head}
The sector embeddings are passed through a GELU dense layer:
\[
\mathbf{Z}_t
=
\operatorname{GELU}\!\left(
\mathbf{B}''_t\mathbf{W}^{g}
\right)
\in\mathbb{R}^{S\times D},
\]
where $\mathbf{W}^{g}\in\mathbb{R}^{D\times D}$. The readout
is \textit{sector-specific}: each sector $s$ has its own weight vector
$\mathbf{w}^{z}_{s}\in\mathbb{R}^{D}$ and bias $b^{z}_{s}$,
reflecting the fact that the dimensions of the shared embedding relevant for
ranking differ across sectors. Writing
$\mathbf{z}_{t,s}=\mathbf{Z}_t[s,:]$, the overweight-side score is:
\[
\hat{z}^{\ow}_{t,s}
=
\mathbf{z}_{t,s}^{\top}\mathbf{w}^{z}_{s}
+
b^{z}_{s},
\qquad s=1,\ldots,S.
\]
Overweight--underweight antisymmetry is enforced by negation:
\[
\hat{\mathbf{z}}^{\act}_t
=
\begin{bmatrix}
\hat{\mathbf{z}}^{\ow}_t\\
-\hat{\mathbf{z}}^{\ow}_t
\end{bmatrix}
\in\mathbb{R}^{2S}.
\]
Since each position is a sector--benchmark spread, overweighting and
underweighting are opposite trades, justifying the sign flip.

The model is trained with the ListMLE objective \cite{xia2008listwise}:
\[
\mathcal{L}_t
=
-\sum_{k=1}^{2S}
\log\!\left(
\frac{
\exp\!\left(\hat{z}^{\act}_t(\pi_t(k))\right)
}{
\sum_{j=k}^{2S}
\exp\!\left(\hat{z}^{\act}_t(\pi_t(j))\right)
}
\right),
\]
where $\pi_t$ orders actions from best to worst by realized rank. Although
the $2S$ scores are determined by $S$ overweight scores through antisymmetry,
applying ListMLE to the full list emphasizes both the strongest overweight
and strongest underweight candidates.
\vspace{5px}
\section{Experiment}
\begin{table*}[t]\small
\centering
% \vspace{-5px}
\caption{Performance comparison of ranking baselines and general methods utilizing a ranking head. Each cell reports the mean and standard deviation (in subscript) across 5 runs.  The best baselines are underlined, and the best results are bolded.}
\label{tab:asset-ranking-results}
\renewcommand{\arraystretch}{0.94}
\begin{tabular}{lccccc}
\toprule
 & Mean Spearman & Pred Acc & NDCG@3 & NDCG@5 & NDCG@11 \\
\midrule
\textbf{Naive Benchmarks} & & & & & \\
Random Ranking & $0.0095_{\pm 0.0143}$ & $0.5029_{\pm 0.0052}$ & $0.5219_{\pm 0.0067}$ & $0.5422_{\pm 0.0068}$ & $0.6142_{\pm 0.0049}$ \\
12-Month Momentum & $-0.0495_{\pm 0.0000}$ & $0.4813_{\pm 0.0000}$ & $0.5097_{\pm 0.0000}$ & $0.5231_{\pm 0.0000}$ & $0.5901_{\pm 0.0000}$ \\
 & & & & & \\
\textbf{Asset Ranking in Finance} & & & & & \\
RankNet \cite{song2017stock}
    & $-0.0243_{\pm 0.0783}$ & $0.4913_{\pm 0.0275}$ & $0.5136_{\pm 0.0251}$ & $0.5311_{\pm 0.0263}$ & $0.6007_{\pm 0.0325}$ \\
ListNet \cite{song2017stock}
    & $-0.0244_{\pm 0.0782}$ & $0.4912_{\pm 0.0274}$ & $0.5137_{\pm 0.0251}$ & $0.5309_{\pm 0.0264}$ & $0.6007_{\pm 0.0325}$ \\
Cross-sectional LTR \cite{poh2020building}
    & $0.0080_{\pm 0.0061}$ & $0.5031_{\pm 0.0021}$ & $0.5220_{\pm 0.0021}$ & $0.5423_{\pm 0.0018}$ & $0.6136_{\pm 0.0023}$ \\
Listwise Node2Vec \cite{saha2021stock}
    & $0.0125_{\pm 0.0116}$ & $0.5045_{\pm 0.0040}$ & $0.5336_{\pm 0.0067}$ & $0.5494_{\pm 0.0054}$ & $0.6172_{\pm 0.0046}$ \\
Long-Short Listwise LTR \cite{zhang2022constructing}
    & $0.0476_{\pm 0.0292}$ & $0.5161_{\pm 0.0099}$ & $0.5370_{\pm 0.0221}$ & $0.5566_{\pm 0.0190}$ & $0.6302_{\pm 0.0100}$ \\
SABER \cite{kouloumpris2024saber}
    & $-0.0818_{\pm 0.0012}$ & $0.4708_{\pm 0.0004}$ & $0.4991_{\pm 0.0004}$ & $0.5124_{\pm 0.0004}$ & $0.5775_{\pm 0.0001}$ \\
 & & & & & \\
\textbf{Asset Ranking in AI} & & & & & \\
RSR \cite{feng2019temporal}
    & $0.0240_{\pm 0.0073}$ & $0.5085_{\pm 0.0028}$ & $0.5307_{\pm 0.0031}$ & $0.5501_{\pm 0.0028}$ & $0.6190_{\pm 0.0027}$ \\
STHAN-SR \cite{sawhney2021stock}
    & $0.0035_{\pm 0.0287}$ & $0.4973_{\pm 0.0162}$ & $0.5116_{\pm 0.0106}$ & $0.5350_{\pm 0.0115}$ & $0.6112_{\pm 0.0104}$ \\
RT-GCN \cite{zheng2023relational}
    & $0.0062_{\pm 0.0187}$ & $0.5015_{\pm 0.0064}$ & $0.5194_{\pm 0.0073}$ & $0.5399_{\pm 0.0077}$ & $0.6134_{\pm 0.0068}$ \\
ADB-TRM \cite{chen2024automatic}
    & $0.0094_{\pm 0.0064}$ & $0.5028_{\pm 0.0022}$ & $0.5248_{\pm 0.0046}$ & $0.5421_{\pm 0.0040}$ & $0.6169_{\pm 0.0023}$ \\
PDU \cite{huang2025progressive}
    & $0.0088_{\pm 0.0188}$ & $0.5032_{\pm 0.0065}$ & $0.5245_{\pm 0.0075}$ & $0.5439_{\pm 0.0081}$ & $0.6138_{\pm 0.0070}$ \\
 & & & & & \\
\textbf{ML + Ranking Head} & & & & & \\
OLS        & $-0.0001_{\pm 0.0000}$ & $0.4995_{\pm 0.0000}$ & $0.5152_{\pm 0.0000}$ & $0.5374_{\pm 0.0000}$ & $0.6104_{\pm 0.0000}$ \\
ElasticNet \cite{zou2005regularization} & $0.0322_{\pm 0.0000}$ & $0.5112_{\pm 0.0000}$ & $0.5421_{\pm 0.0000}$ & $0.5579_{\pm 0.0000}$ & $0.6221_{\pm 0.0000}$ \\
XGBoost \cite{chen2016xgboost}   & $0.0391_{\pm 0.0047}$ & $0.5139_{\pm 0.0015}$ & $0.5374_{\pm 0.0022}$ & $0.5570_{\pm 0.0023}$ & $0.6251_{\pm 0.0027}$ \\
GNN \cite{kipf2016semi}       & $0.0463_{\pm 0.0121}$ & $0.5165_{\pm 0.0039}$ & $0.5440_{\pm 0.0039}$ & $0.5611_{\pm 0.0040}$ & $0.6290_{\pm 0.0049}$ \\
Diffusion \cite{ho2020denoising} & $\underline{0.0708}_{\pm 0.0178}$ & $\underline{0.5246}_{\pm 0.0063}$ & $\underline{0.5491}_{\pm 0.0107}$ & $\underline{0.5702}_{\pm 0.0092}$ & $\underline{0.6377}_{\pm 0.0072}$ \\
 & & & & & \\
\textbf{Our Model} & & & & & \\
\textsc{RegimeRank} & $\mathbf{0.0997}_{\pm 0.0163}$ & $\mathbf{0.5350}_{\pm 0.0062}$ & $\mathbf{0.5708}_{\pm 0.0118}$ & $\mathbf{0.5870}_{\pm 0.0100}$ & $\mathbf{0.6498}_{\pm 0.0063}$ \\
\bottomrule
\end{tabular}
% \vspace{-5px}
\vspace{7px}
\end{table*}
\subsection{Dataset}
We source \emph{weekly} sector-level fundamental, technical and
shared macro data from Bloomberg, covering 11 sectors with
49 sector-specific features and 156 shared macro
features. Available dates range from January 2002 to December
2024; as the label requires a full forward year, the last usable
observation date is December 2023. Fundamental data is point-in-time.
The target ranking is constructed from the 1-year forward Sharpe, computed from weekly returns over the following 52 weeks.

%All
%feature transformations are fitted on training data only within
%each fold.

\subsubsection{Combinatorial Purged Cross-Validation}
To account for the overlapping forward labels and prevent information leakage,
we adopt combinatorial purged cross-validation (CPCV) \cite{joubert2024enhanced, bailey2017probability}.

\begin{figure}[h]
\centering
\includegraphics[width=\linewidth]{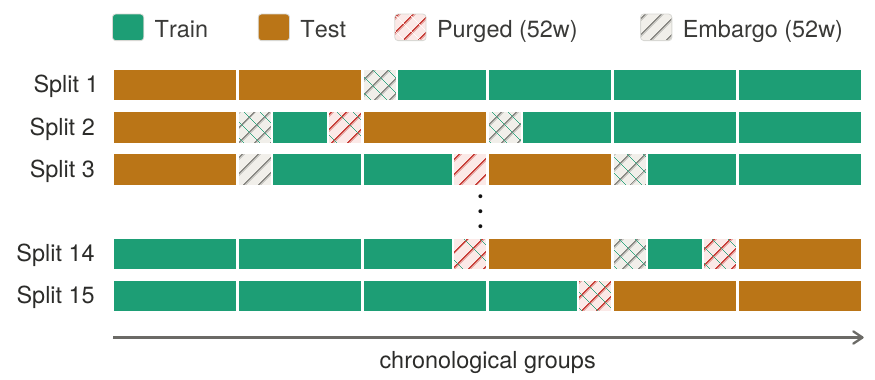}
\caption{We adopt combinatorial purged cross-validation, which is often used in financial machine learning \cite{joubert2024enhanced, bailey2017probability}.}
\label{fig:cpcv}
\end{figure}

The complete sample is first divided chronologically into six groups; see Figure \ref{fig:cpcv}. For each split, two groups are selected as the test set, while the remaining groups are used for training. This gives a total of $\binom{6}{2}=15$ train-test combinations.

\paragraph{Purge and embargo.}
The purge and embargo lengths arise from two distinct channels: the
forward label horizon and the backward feature window. Preceding each
test group we purge 52 weeks, matching the label span exactly---a
training observation at $t$ carries a label realised over $[t, t+52]$---%
thereby removing in full the channel through which target information
propagates directly. Following each test group we set an embargo of 52 weeks. 
% in line with embargo fraction typically recommended for CPCV~\citep{lopezdeprado2018}. 
This addresses a weaker
channel: a training observation placed shortly afterwards draws part of
its \emph{inputs} from realisations inside the test period for calculating temporal transformations such as volatility, trend, \etc
% \footnote{We note that 7 weeks is significantly shorter than our longest
% trailing transformation window, so this channel is attenuated rather than
% eliminated; a fully clean separation would theoretically require an embargo of 156
% weeks at each of the 15 fold combinations, discarding much of an already
% modest weekly sample. Two considerations limit the practical impact:
% (1) the longest window is used for z-scoring, which enters only through
% location and scale, so any single overlapping observation carries weight
% of order $1/L$ ($L \approx 156$ weeks) and the statistic is dominated by
% the contemporaneous value; (2) the residual dependence takes the form of
% correlation between adjacent training and test features rather than
% access to test-period targets, so its likely effect is a mildly
% optimistic variance estimate rather than a lookahead bias.}.

Because we use weekly observations with a 52-week forward label, consecutive training examples exhibit overlap in their target windows. Specifically, the full sample of roughly 1{,}200 weeks yields only about 23 non-overlapping label periods. Consequently, each training fold contains roughly 13--15 independent periods, and each test fold contains fewer than 8. The cross-section offers limited additional independence, as sector returns are highly contemporaneously correlated and the regime-conditioning component encounters only a modest number of \emph{distinct} macroeconomic regimes over 2002--2024. This constraint on independent samples is inherent to long-horizon financial prediction rather than specific to our setup. It directly motivates our combinatorial design: while a single hold-out split would rely on a fragile number of independent periods, evaluating across all 15 combinations exposes each model to a substantially broader range of market conditions.

\begin{comment}
With weekly observations and a 52-week forward label, consecutive training
examples share the large majority of their target window: the
full sample of roughly 1{,}200 weeks corresponds to
approximately 23 non-overlapping label periods, each training
fold of 682--793 weeks to roughly 13--15, and each test fold of
approximately 400 weeks to fewer than 8. The cross-section
provides limited additional independence, as sector returns are
strongly contemporaneously correlated, and the
regime-conditioning component sees only a modest number of
\emph{distinct} macroeconomic regimes over 2002--2024. This
constraint is inherent to long-horizon financial prediction
rather than specific to our setup, and it motivates the
combinatorial design: a single hold-out split would rest on a
small number of independent periods, whereas evaluating over all
15 combinations exposes each model to a wider range of market
conditions.
\end{comment}

% \vspace{-3px}
\subsection{Baselines}
We benchmark \textsc{RegimeRank} against four groups of baselines. 

The first group, \textbf{Naive Benchmarks}, contains a random ranking model, which establishes the chance level for each performance metric, and a 12-month momentum rule, which ranks the long and short sector actions based solely on their trailing 12-month performance. The next two groups contain state-of-the-art asset ranking models discussed in Table \ref{tab:asset-ranking-results}. 

For \textbf{Asset Ranking in Finance}, we have: RankNet and ListNet \cite{song2017stock}, Cross-sectional Learning to Rank (LTR) \cite{poh2020building}, Listwise Node2Vec \cite{saha2021stock}, Long-Short Listwise LTR \cite{zhang2022constructing}, and Stochastic-Aware Bootstrap Ensemble Ranking (SABER) \cite{kouloumpris2024saber}. 

For \textbf{Asset Ranking in AI}, we have: Relational Stock Ranking (RSR) \cite{feng2019temporal}, Spatio Temporal
Hypergraph Attention Network for Stock Ranking (STHAN-SR) \cite{sawhney2021stock}, Relational Temporal Graph Convolutional Network (RT-GCN) \cite{zheng2023relational}, Automatic De-Biased Temporal-Relational Modeling (ADB-TRM) \cite{chen2024automatic}, and Progressive
Dependency representation learning with Uncertain risk contrasting (PDU) \cite{huang2025progressive}. 

In the final group, we compare with general machine learning
methods fitted with the same listwise ranking head used in our
model. These are: OLS, ElasticNet \cite{zou2005regularization},
XGBoost \cite{chen2016xgboost}, a
graph neural network (GNN) \cite{kipf2016semi}, and a Diffusion
model \cite{ho2020denoising}.

% \vspace{-3px}
\subsection{Experimental Settings}
\begin{table*}[t]\small
\centering
% \vspace{-5px}
\caption{Ablation study. Each cell reports the mean and standard deviation (in subscript) across 5 runs.}
% \vspace{-7px}
\label{tab:ablation}
\renewcommand{\arraystretch}{1.0}
\begin{tabular}{lccccc}
\toprule
 & Mean Spearman & Pred Acc & NDCG@3 & NDCG@5 & NDCG@11 \\
\midrule
w/o Latent Market States      & $0.0662_{\pm 0.0301}$ & $0.5232_{\pm 0.0108}$ & $0.5530_{\pm 0.0144}$ & $0.5704_{\pm 0.0145}$ & $0.6365_{\pm 0.0119}$ \\
w/o Regime Conditioning       & $0.0870_{\pm 0.0135}$ & $0.5300_{\pm 0.0044}$ & $0.5593_{\pm 0.0058}$ & $0.5771_{\pm 0.0058}$ & $0.6449_{\pm 0.0056}$ \\
w/o Sector Self-Attention & $0.0358_{\pm 0.0186}$ & $0.5125_{\pm 0.0070}$ & $0.5372_{\pm 0.0104}$ & $0.5560_{\pm 0.0096}$ & $0.6240_{\pm 0.0071}$ \\
\textsc{RegimeRank} (Ours)      & $\mathbf{0.0997}_{\pm 0.0163}$ & $\mathbf{0.5350}_{\pm 0.0062}$ & $\mathbf{0.5708}_{\pm 0.0118}$ & $\mathbf{0.5870}_{\pm 0.0100}$ & $\mathbf{0.6498}_{\pm 0.0063}$ \\
\bottomrule
\end{tabular}
% \vspace{-7px}
% \vspace{3px}
\end{table*}
The Adam optimizer is used to optimize the model, with an initial learning rate of $1\times10^{-3}$ and a weight decay of
$1\times10^{-4}$. The batch size is set to 64.
% , and the model is trained for 80 epochs. 
% \textbf{[TODO: resolve early stopping ---
% either "Within each training fold, the final contiguous block is
% held out for validation, with the same purge and embargo applied
% at its boundaries; training stops when validation performance
% ceases to improve" or "All models are trained for a fixed budget
% of 80 epochs, fixed in advance and applied uniformly across
% methods".]} 
All models are trained for a fixed budget of 80 epochs, applied uniformly across methods.
We set the learnt embedding dimension to $D=128$,
with a feed-forward hidden dimension of 512 and a dropout rate of
0.1.

For both the regime-conditioning cross-attention and cross-sector self-attention components, we use 4 attention heads. The cross-sector self-attention block is also repeated across $L=2$ layers.

\subsection{Evaluation Metrics}
We evaluate models over multiple out-of-sample metrics: \textbf{Pairwise accuracy} measures the ratio of ranking pairs that are ordered correctly by the model; \textbf{Spearman's rank correlation} \cite{spearman1961proof} evaluates the correlation between the set of predicted and realized rankings; \textbf{Normalized Discounted Cumulative Gain (NDCG)} \cite{jarvelin2002cumulated} calculates the relevance of the results discounted by their position in the list. Specifically, we use NDCG@3, NDCG@5 and NDCG@11. Higher values indicate better performance for all metrics.

All experiments are repeated five times and the mean and
standard deviation are reported. Reported standard deviations
are computed across the five training runs and reflect
run-to-run stability. As the 15 test combinations overlap by
construction, their dispersion does not constitute an
independent error bar. 

% Future additions to this paper:

%The fourth group contains \textbf{heuristic baselines} that require
%no learning: \textbf{Random}, which ranks sectors at random and
%establishes the floor for each metric; \textbf{Persistence}, which
%ranks by trailing 52-week realised Sharpe ratio; and
%\textbf{Momentum}, which ranks by trailing 12-month return skipping
%the most recent month. These provide a reference point against
%which the value of a learned ranking can be assessed.
% \vspace{-15px}
\section{Results} 
\subsection{Performance Comparison.}
Table~\ref{tab:asset-ranking-results} reports the ranking results. We can observe the following:
\begin{itemize}[leftmargin=*, itemsep=0.3em]
% \begin{itemize}[leftmargin=*, itemsep=0.3em]
\item Overall, \textsc{RegimeRank} matches or exceeds well-established ranking baselines, suggesting that the constraints of interpretability and expert-informed structure cost little in ranking accuracy.
\item The naive benchmarks calibrate the table. Random ranking pins each metric at its chance level (accuracy $\approx 0.50$, NDCG@3 $\approx 0.52$), revealing that several published baselines sit at or below chance. The 12-month momentum rule falls below random, indicating that trailing one-year sector performance mean-reverts at this horizon rather than persisting.
\item Among the finance baselines, the strongest add structure to the plain cross-sectional objective: Listwise Node2Vec encodes asset relationships through node embeddings, and Long-Short Listwise LTR optimizes a relative rather than absolute target. Both improve on Cross-sectional LTR: the first suggesting a sector's rank depends on other sectors, the second that a relative target is the more informative signal. Our architecture combines both.
\item The AI ranking models underperform the simpler baselines and exhibit high variance. RSR is the strongest of this group, benefiting from temporal-relational modeling, whereas the more complex graph and hypergraph models (\ie STHAN-SR, RT-GCN) collapse to near-zero mean Spearman with high variance. With the smaller sector-level universe, these high-capacity relational models appear more prone to overfitting.
\item ElasticNet and XGBoost perform well with a ranking head, consistent with the paucity of independent observations at our horizon: a one-year forward target over two decades yields few effectively independent outcomes, which favours parsimonious models. The strongest baseline, Diffusion, leads on every metric; its denoising formulation handles the noise inherent in financial signals, while GNN, which learns sector relations, follows close behind.
\item \textsc{RegimeRank} achieves the best result on every metric. It improves mean Spearman correlation by $40.8\%$ over the strongest baseline and lifts pairwise accuracy by $2.0\%$ (relative). Notably, its largest relative gains fall on the top-ranked NDCG (NDCG@3/@5), the metrics most relevant to portfolio positioning.
\item SABER scores poorly here, perhaps because it optimizes a risk-shrunk portfolio objective rather than a ranking one, a gap that an economic evaluation might close or reverse.
\end{itemize}

% \vspace{-15px}
% \vspace{-5px}
\subsection{Ablation Study.}
Table~\ref{tab:ablation} reports the contribution of each component, measured by removing it from the full model. We observe the following:

\begin{figure}[t]
  \centering
  \includegraphics[width=\columnwidth]{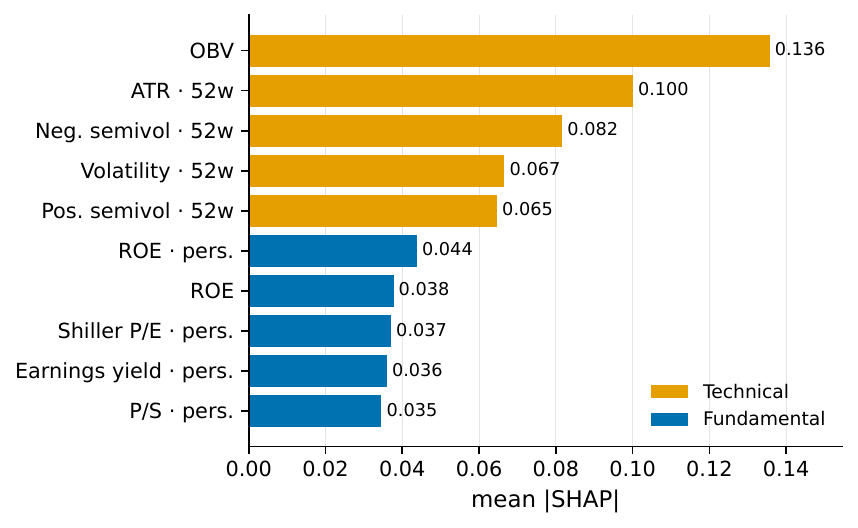}
  \caption{Top-5 attributions for each feature type.}
  \label{fig:pillar}
  \par\vspace{\floatsep}
  \includegraphics[width=0.667\columnwidth]{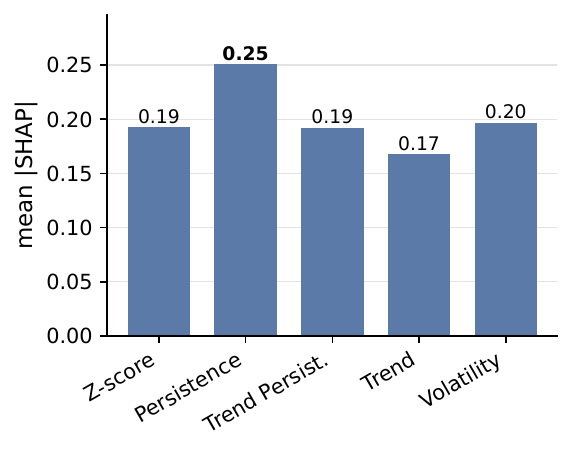}
  \caption{Attribution by temporal transformation.}
  \label{fig:transform}
\end{figure}

\begin{figure*}[t]
  \centering
  \begin{minipage}[b]{0.70\textwidth}
    \centering
    \includegraphics[width=\linewidth]{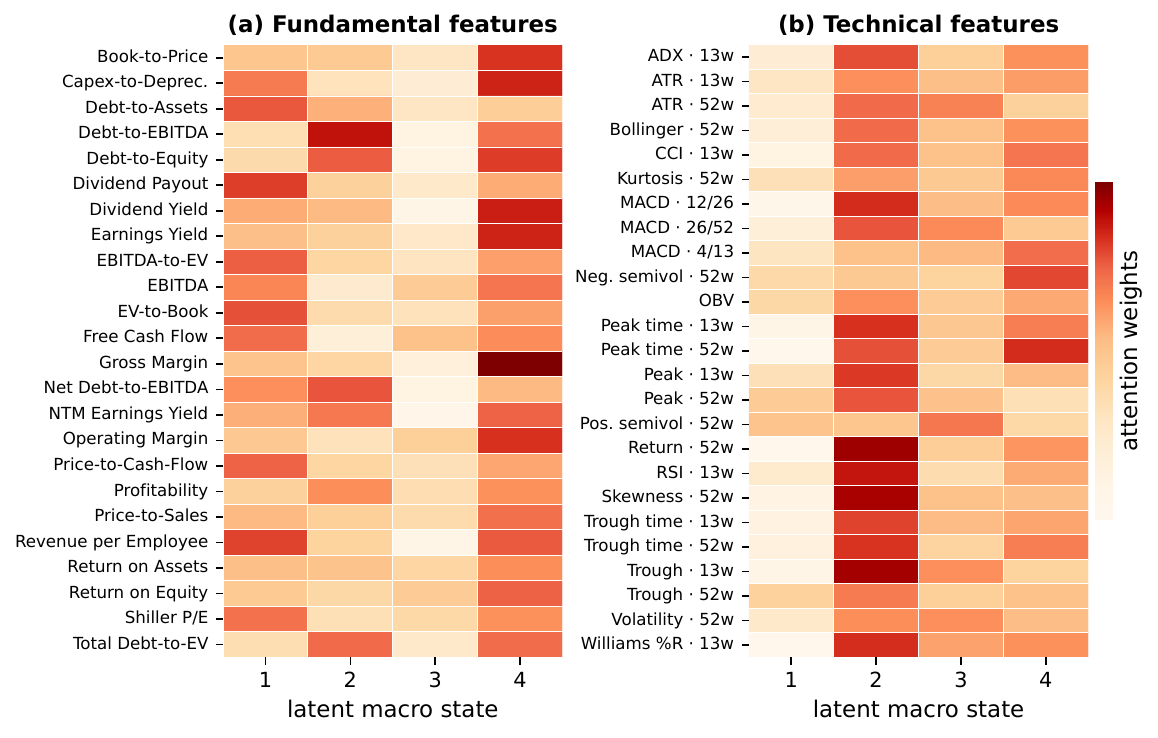}
    % \vspace{-20px}
    \caption{Attention weights for (a) fundamental and (b) technical features.}
    \label{fig:conditioning}
  \end{minipage}\hfill
  \begin{minipage}[b]{0.30\textwidth}
    \centering
    \includegraphics[width=0.942\linewidth]{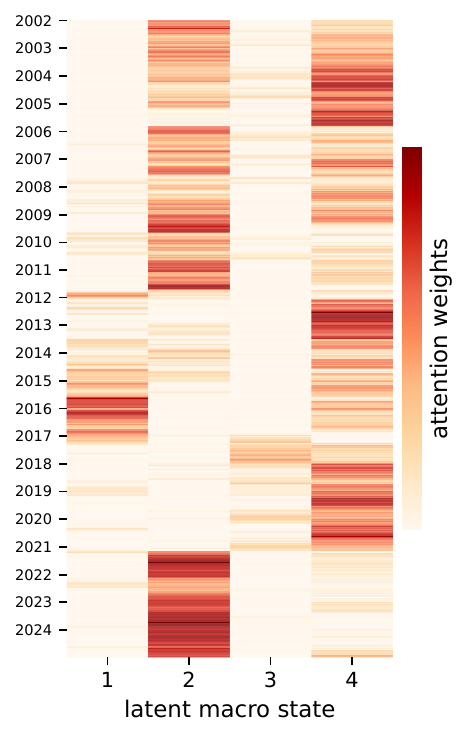}
    % \vspace{-20px}
    \caption{Attention weights over time.}
    \label{fig:regime}
  \end{minipage}
  % \vspace{-8px}
\end{figure*}

\begin{itemize}[leftmargin=*, itemsep=0.3em]
% \begin{itemize}[leftmargin=*, itemsep=0.3em]
\item Sector self-attention matters most, with the largest drop on every metric: sector interdependencies are central to the task.
\item Latent market states come second: removing them lowers the mean and raises the run-to-run variance on every metric. Conditioning on raw macro features directly performs worse than removing regime conditioning altogether, indicating that the latent-state bottleneck is what makes macro information usable.
\item Regime conditioning contributes the smallest yet consistent increment: the prevailing macro state still adds information that the sector-level features alone do not carry.
\end{itemize}

% \vspace{-5px}
\subsection{Interpretability Study.}
\noindent\textbf{Feature Attribution.}
Next, we interpret the feature and transform importance through their SHAP values $\boldsymbol{\Phi}_t$. We can observe:

\begin{itemize}[leftmargin=*, itemsep=0.3em]
% \begin{itemize}[leftmargin=*, itemsep=0.3em]
\item From Figure \ref{fig:pillar}, the five largest attributions are all technical signals, each exceeding the strongest fundamental. For a one-year forward Sharpe target, price and volatility dynamics carry more of the ranking signal than valuation fundamentals --- OBV (on-balance volume) leads, ahead of 52-week ATR (average true range), semivolatility and volatility.
\item From Figure \ref{fig:transform}, \textit{persistence} is the most influential temporal transformation, ahead of $z$-score, trend-persistence, volatility and trend. The pattern also holds within the fundamental pillar, where the most influential features come from the persistence channel rather than any other transformation.
\item This supports the autocorrelation premise behind our transformation-projection layer: how strongly a signal continues appears more informative for cross-sectional ranking than its level or direction.
\end{itemize}

\noindent\textbf{Macro Regime Conditioning.}
We interpret the latent macro states through their cross-attention weights $\Omega^{(s)}_t$. We observe:

\begin{itemize}[leftmargin=*, itemsep=0.3em]
% \begin{itemize}[leftmargin=*, itemsep=0.3em]
\item From Figure \ref{fig:conditioning}, feature importance varies with the learnt regimes: technical features weigh most heavily under latent state 2, fundamentals under latent state 4. The active states thus separate into price-driven and valuation-driven regimes, recovering a regime-dependence of signal informativeness that a standard cross-sectional model discards by construction.
\item From Figure \ref{fig:regime}, these states align with recognisable market periods. State 2 dominates the volatile 2002--2010 window, peaking around the 2008 \textbf{Global Financial Crisis}, and resurges over 2021--2024; state 4 activates in the calmer intervening years. The remaining states stay largely dormant. The model therefore reweights features as conditions change, leaning on technical signals in stressed markets and on fundamentals when volatility subsides.
\end{itemize}

\begin{figure}[t]
  \centering
  % \vspace{-5px}
  \includegraphics[width=\columnwidth]{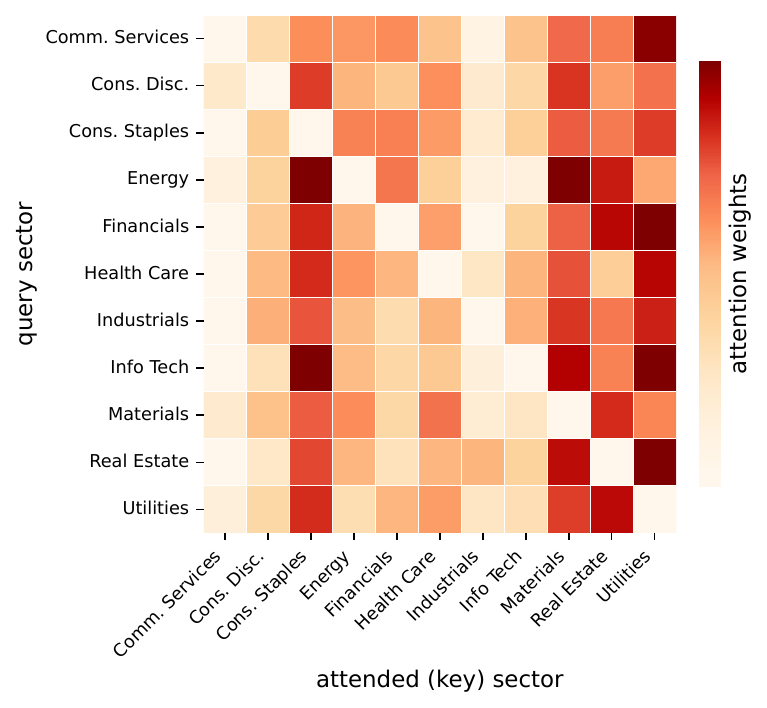}
  % \vspace{-9px}
  \caption{Cross-sector self-attention weights.}
  \label{fig:sector}
  % \vspace{-20px}
  \vspace{2px}
\end{figure}

\noindent\textbf{Cross-Sector Interdependencies.}
We also study the sector interdependencies using their self-attention weights $\Gamma_t$. We can see:

\begin{itemize}[leftmargin=*, itemsep=0.3em]
% \begin{itemize}[leftmargin=*, itemsep=0.3em]
\item In Figure \ref{fig:sector}, the sample-averaged attention concentrates on a few sectors: nearly every query sector attends strongly to Real Estate, Materials, Consumer Staples and Utilities.
\item The diagonal is weak, indicating that sectors draw on \emph{other} sectors' signals rather than reinforcing their own, whose information is already preserved through the residual connection $\mathbf{B}_t$.
\item The learnt structure is economically coherent, with strong attention between cyclical sectors (\eg Energy and Materials) and among the rate-sensitive Real Estate, Financials and Utilities, consistent with a risk-based view of the cross-section \cite{kelly2025artificial}.
\end{itemize}

% \vspace{-4px}
\section{Conclusion} 

We presented \textsc{RegimeRank}, an interpretable cross-sectional ranking model for long-short sector allocation that recovers temporal dynamics through signal transformation projections and conditions features on latent macro regimes via cross-attention. Across our experiments it ranks ahead of the baselines we consider on several ranking metrics, while its SHAP values and attention weights expose attributions over feature transformations, macro regimes, and sector interdependencies that portfolio managers can inspect.
Overall, these findings highlight the complementary roles of temporal signal representations, macroeconomic conditioning, and cross-sector interactions in producing interpretable sector rankings.

\noindent\textbf{Future Work.}
Three directions follow. First, we plan to evaluate economic value directly, translating the top-ranked over- and underweights into portfolio positions and measuring the fund performance they induce, net of turnover and trading costs. Second, we plan to add a sentiment pillar, which the architecture accommodates as-is, to test whether news and analyst signals rank sectors beyond fundamentals and technicals. Third, we plan to translate the structured attributions into verbal explanations through large language models, producing rationales PMs can read directly.

\clearpage
\bibliographystyle{ACM-Reference-Format}
\bibliography{reference_long}
% \bibliography{reference}

\appendix

\end{document}